# A Disposable Soft Magnetic Ribbon Based Sensor for Corrosion Monitoring

Izabella Berman[1,2], Jagannath Devkota[1], Kee Young Hwang[1], and Manh-Huong Phan[1,*]

[1] *The Laboratory for Advanced Materials and Sensors, Department of Physics, University of South Florida, Tampa, FL 33620, USA*

[2] *Department of Physics, Beloit College, Beloit, WI, 5351, USA*

We present a new approach for real-time monitoring of chemical corrosion based on the radio-frequency (RF) impedance technology and soft ferromagnetic ribbons. The impedance ($Z$) of a commercial METGLAS® 2714A ribbon was measured in real time for 5 μl of drop-casted $HNO_3$ of various concentrations. Variations in the concentration of the drop-casted acid were assessed by considering the difference ($\Delta$) in $Z$ with and without the acid treatment. We found a large and linear increase in $\Delta Z$ and a large linear decrease in measurement time with the acid concentration, which are ideal for developing disposable chemical sensors for strength estimation of corrosive chemicals and for monitoring of a time-dependent chemical corrosion process. Since the ribbon used is commercially available at low cost and the measurement system is quick and low power consuming, the proposed sensor can be used as an easy, quick, and low-cost chemical probe in industries and environmental hazard management purposes.

*Corresponding author: phanm@usf.edu

# 1. Introduction

Development of a quick, cost effective, and reliable technique to estimate the concentration of corrosive chemicals has been of technological interest for safety in industries and the environment for many years [1-15]. Numerous different sensors have been established to achieve this goal such as titanium oxide based semiconductor sensors used to detect formaldehyde in the air [6], and enzymatic conductometric sensors commonly used to detect the amount of pollution in water samples [3,7]. Each different type of sensor carries with it advantages and disadvantages which are useful for different situations. Capacitive sensors are praised for their easy fabrication and cost effective construction, yet can provide trouble since their sensitivity depends on obtaining the proper thickness of the original sensing layer [7]. Metal oxide semiconductor gas sensors are highly sensitive, simple, and have the possibility to be portable and low cost yet are only applicable to chemicals in the gas phase [1,6,11]. Throughout the variety of chemical sensors available today, there is an evident lack of chemical sensors to distinguish between different concentrations and types of corrosive chemicals in liquid form, which are often used in industries such as the steel industry, on a daily basis. Here, a novel sensor is proposed for this issue which utilizes impedimetric sensing benefits of electrochemical sensors [2,8-10,12], while also taking advantage of the inductive impedance benefits of the recently popular giant magneto-impedance (GMI) based chemical sensors [13-17].

Electrochemical impedimetric sensors exhibit many advantages that have aided in their importance in today's society [2,8-10,12]. The main asset of these sensors is their ability to provide information on different aspects of electrochemical cells such as the ohmic resistance, capacitive information at the electrode/solution interface, and the rate constant of the reactions involved [2,8,11]. Through this information it is possible to determine many characteristics about a specific analyte in question such as the presence of bacteria in water samples [12], the concentration of humidity in the air [2], and the presence of certain proteins within the body [10-12]. Unfortunately, many impedimetric electrochemical sensors require external chemicals aside from those being detected in order to function, such as the enzymatic conductometric sensors which utilize enzymatic reactions that produce conductivity modifying ionic species of the reaction solution [3]. These extra chemicals can increase the complexity of use and cost of the sensors. Recently, another form of impedance based sensors referred to as GMI sensors has come into importance for their utilization of inductive impedance using soft ferromagnetic ribbons and external magnetic fields [13-17]. GMI sensors are praised for their low power consumption, high sensitivity and thermal stability, which have yielded accurate detections for different concentrations of corrosive chemicals such as $HNO_3$ [16], but similar to the electrochemical impedimetric sensors, require an external magnetic field which adds to the complexity in development and use. The novel sensor created utilizes advantages of both impedimetric electrochemical sensors and GMI sensors by exploring the benefits of the GMI sensors without the added external magnetic field. Such a sensor that utilizes the ac impedance, $Z$, of a magnetic ribbon of half thickness $t$, resistivity $\rho$, and magnetic permeability $\mu_T$, at a frequency $f$, and skin depth, $\delta$ has the potential to improve safety in chemical using industries. The impedance of the previously mentioned class of sensors is given by [18]

$$Z = R_{dc} jkt \coth(jkt)$$

where

$$k = \frac{1+j}{\delta}$$

and

$$\delta = \sqrt{\frac{\rho}{\pi \mu_T f}}$$

$\rho$ is given by

$$\rho = R\left(\frac{A}{l}\right)$$

$R$ is the resistance of the ribbon, $A$, the cross-sectional area, and $l$ the length.

For inductive materials, as the frequency is increased, the skin depth decreases, which leads to an increase in the impedance of the system. When treated with different concentrations of corrosive chemicals, parts of the ribbon are etched away and holes accompanied by metal oxide on the surface of the ribbon form [16,17]. This change in surface topography of the ribbon decreases the magnetic permeability, as magnetic moments are removed from the surface as the ribbon is etched. According to the above equations, if it is assumed that the frequency of oscillation remains constant after acid treatment, this would lead to an increase in the skin depth. Plugging the skin depth into the $Z$ expression, it can be noted that the coth(jkt) section will increase along with the $R_{dc}$ section. This leads to an overall increase in the impedance which allows the change in impedance to be analyzed to give the concentration of the corrosive chemical.

In this study, the sensitivity, defined by the change in impedance of soft ferromagnetic ribbons, is studied using frequencies ranging from 0.2 MHz to 7 MHz and concentrations of $HNO_3$ ranging from 0.9M to 7.4M. In addition, the change in morphology and magnetic properties of the ribbon are studied using Scanning Electron Microscopy (SEM) imaging, X-Ray Diffraction (XRD) data, and Physical Property Measurement System (PPMS) data. The proposed sensor would have low power consumption, be cost effective, disposable after use, and would not require the complexities of external magnetic fields or additional chemicals. By the creation of a simple to use corrosive chemical sensor, emergencies such as chemical spills in industries have the possibility to be dealt with in safer manners, and the safety of workers in chemical using industries can greatly increase.

## 2. Materials and Methods

### 2.1. Sensor design and fabrication

Soft ferromagnetic amorphous ribbons of composition $Co_{65}Fe_4Ni_2Si_{15}B_{14}$ (METGLAS 2714A) were cut into 3 mm wide strips so that the magnetic domains are transverse to their length. The impedance of the strips was measured by a four point measurement technique over the length of 15 or 20 mm using an HP4192 impedance analyzer for a test current of 1.36 mA. The electric contacts were made using copper wires and silver paint and the whole measurement system was monitored by LabVIEW program. A detail of the experimental setup is shown in Figure 1.

### 2.2. Tests and analysis of chemical corrosion

The effect of the $HNO_3$ strength on the impedance was accessed by their real time measurement by drop-casting a certain volume (~10 uL) of $HNO_3$ with concentrations 7.4 Molar, 5.5 Molar, 3.8 Molar, 1.8 Molar, and 0.9 Molar. All the measurements were repeated for the test current of frequencies 0.2 MHz, 2 MHz, 5 MHz, and 7 MHz to determine the optimal frequency range and acid concentration the sensor could function at. Ten trials were conducted for each frequency and acid concentration, using different identical ribbons each time, and the average of the data was then used for analysis. From the data stated, graphs containing the change in impedance as a function of acid concentration, and change in impedance as a function of time could be obtained and analyzed.

### 2.3. Topological, structural and magnetic characterization

The ribbons were then subjected to the acid concentrations previously listed for ten minutes each, and studied using Scanning Electron Microscopy (SEM) imaging to better understand the topological effect the acid would have on the ribbons. Imaging of the ribbons was done both when wiping the acid off of the ribbons after the designated ten minutes, and with keeping the acid on after the ten minutes. Along with the SEM imaging, the ribbons were also studied using X-ray diffraction methods (XRD) after the acid treatment to ensure that the change in the conductivity observed in the ribbons as a function of acid concentration, was not due to structural changes in the ribbons.

Aside from the structural, topological, and electronic changes measured in the ribbons, magnetic changes of the ribbons as a function of acid concentration were also measured using a vibrating sample magnetometer (VSM) probe equipped within the physical property measurement system (PPMS) from Quantum Design. Ribbons were subjected to the previously stated concentrations of $HNO_3$, cleaned using acetone, weighed before and after acid treatment, and then subjected to a magnetic field of varying strength from -300 Oe to 300 Oe. From this data the hysteresis loops of the ribbons along with the change in saturation magnetization were obtained before and after acid treatment.

## 3. Results and Discussion

Figure 2(a-d) shows the images obtained from the SEM and XRD studies. Since in real time the nitric acid is not wiped off of the ribbon, the results shown are of only the images where the acid was also not wiped off. It can be seen that the surface of the ribbon experienced significant corrosion from the acid, as well as produced a significant amount of metal oxide in response to the acid treatment (Figure 2b,c). This in turn can change the geometry and magnetic properties of the ribbon, leading to a change in impedance. No structural change in the ribbons was observed

by the XRD study (Figure 2d). These results suggested that the change in impedance of the ribbons was most likely a product of a change to the magnetic permeability brought about by the nitric acid's corrosion of the surface, along with a change in the topography of the ribbon, which effects the resistivity. According to the above equations, the frequency range chosen for the ac signal would also affect the sensitivity of the sensor as well. SEM images showed changes to the morphology of the ribbon which decreases the area $A$, thus, assuming the change of $R$ is small compared to the area, decreases the resistivity. It would seem a reasonable assumption for the change in impedance to have dependence on the frequency, geometry, and magnetic permeability.

The results from the PPMS furthered the legitimacy of this assumption by showing a decrease of saturation magnetization ($M_S$) for the ribbon as a function of acid concentration, which can be seen in Figure 3a,b. It can be seen that while saturation magnetization decreased, no obvious effect exists on the magnetic coercivity or anisotropy of the ribbon. The decrease in $M_S$ has been previously reported for acid-treated ribbons, when the acid concentration was increased [16,17].

The frequency dependence was also tested for the impedance based sensor, where the acid concentration used to test the frequency dependence was 7.4 M, due to its large effects on the topography of the ribbon. It was found that the data using frequencies above 2 MHz possessed high amounts of noise, making its reliability questionable. For this reason, the frequency of 0.2 MHz was chosen to conduct the rest of the studies, as it exhibited low noise and accurate data. Our previous GMI study [16] has shown 2 MHz to be the optimal frequency of measurement, and further research with improved noise correction should be conducted to investigate the legitimacy of this frequency when applied to this specific sensor.

Figure 4a-c shows the results of impedance testing done as a function of acid concentration, using the chosen frequency of 0.2 MHz. Figure 4a shows the change in impedance as a function of time for different acid concentrations. It is clearly seen that as the acid concentration increases, the sensitivity of the sensor increases as well, indicating efficient sensing capability. Figure 4b shows the graph of the sensor sensitivity, $\eta$, as a function of acid concentration. In Figure 4b, a large slope is observed for the lower acid concentration, which lessens as the concentration increases, indicating a possible limit to the damage obtained to the ribbon from the acid. The linear fit on the graph indicates the sensor in question would yield accurate and sensitive results for chemical sensing when properly calibrated, as well as function in simple, easy to use manners for all users.

The spacing between the two copper wires which supplied the current to the ribbon was also found to influence the sensitivity of the sensor. Because the reaction time of $HNO_3$ increases with lower concentrations, the acid was provided with more time to expand than for higher concentrations. Due to this occurrence, interference between the wires and acid was observed for separations of less than 20 mm, leading to 20 mm being the ideal separation of the wires. It was also found that the reaction time of the change in impedance also yields important information on the concentration of corrosive chemicals, and shows an almost linear decrease in reaction time as the acid concentration grows (Figure 4c). Therefore, this aspect of the sensor can simultaneously be used as a sensing unit as well.

It is generally known that the ac impedance ($Z$) consists of its real ($R$, ac resistance) and imaginary ($X$, inductance) parts [18]. A fundamental question thus emerges as to how $R$ and $X$ were altered subject to the acid treatment. It can be clearly seen in Figure 5, that the impedance is mostly resistive in this sensor, and thus provides the main contribution to the sensitivity, as was observed from impedance analyzer data. This fact supports further investigation of a more sensitive outcome at higher frequencies than looked at in this study.

An impedance based corrosive chemical sensor to be used with liquid corrosive chemicals is a novel addition to the field of impedance-based sensors [18,19]. Many sensors have been constructed for use in the medical field as biosensors for drug delivery and cancer monitoring [18-20]. Many more have been constructed using impedance spectroscopy for chemical detection in the atmosphere, but usually using non-conductive, non-magnetic materials. GMI effects have been studied with applications to corrosive chemical sensors [13-17], and while GMI techniques are known for their sensitivity, they also require an external magnetic field and often a reference probe. The design of this new chemical sensor is easy to use and construct, and has the potential to be more cost effective and energy efficient than the GMI chemical sensors due to its ability to function with no external magnetic field or chemicals, while also operating without any form of reference probe. The exploitation of GMI sensitivity in conjunction with the simplicity of many electrochemical sensors has brought about a novel impedance-based sensor which can be used to improve safety in many chemical using industries.

## 4. Conclusions

We have proposed a new approach based on radio-frequency impedance technology and soft ferromagnetic ribbons for real-time monitoring of chemical corrosion. A large change in the ac impedance with the acid concentration has been found, demonstrating the possibility of developing disposable chemical sensors for strength estimation of corrosive chemicals such as $HNO_3$. Since the ribbon used is commercially available at low cost and the measurement system is quick and low power consuming, the proposed sensor can be used as an easy, quick, and low-cost chemical probe in industries and environmental safety purposes.

**Acknowledgements**

The authors acknowledge support from the CAS-USF grant. I.B acknowledges also support from the USF-NSF REU program (Grant number: DMR-1263066). Metglas/Hitachi Metals America was acknowledged for providing the ribbons. The authors also thank Prof. Hari Srikanth for his useful discussions.

**Figure captions**

**Figure 1.** Schematic of the impedance-based measurement system using a soft ferromagnetic ribbon as a chemical sensing element. Photograph of a four probe measurement using the ribbon.

**Figure 2.** SEM images of the (a) untreated and (b,c) acid-treated (7.4 Molar $HNO_3$) ribbons. XRD patterns of the untreated and acid-treated (7.4 Molar $HNO_3$) ribbons with reference to the sample holder.

**Figure 3.** (a) Magnetic hysteresis (M-H) loops taken at 300 K for the untreated and acid-treated ribbons for different concentrations; (b) Saturation magnetization is plotted as a function of acid concentration.

**Figure 4.** (a) Changes in the ac impedance (*Z*) at different acid concentrations as a function of measurement time at a fixed frequency of 0.2 MHz. As the acid concentration increases, a stronger change in impedance can be observed; (b) shows the sensor sensitivity, $\eta$, as a function of acid concentration. The linear fit displayed on the graph implies reliable and efficient sensing capability; (c) shows the reaction time as a function of acid concentration, indicating another potential sensing source from the sensor.

**Figure 5.** The impedance, *Z*, and its components, the resistance, (*R*), and reactance, (*X*) as a function of measurement time during acid treatment. It can be observed that at the chosen frequency of 0.2 MHz, the main contribution to the sensitivity comes from the resistance.

**Figure 1**

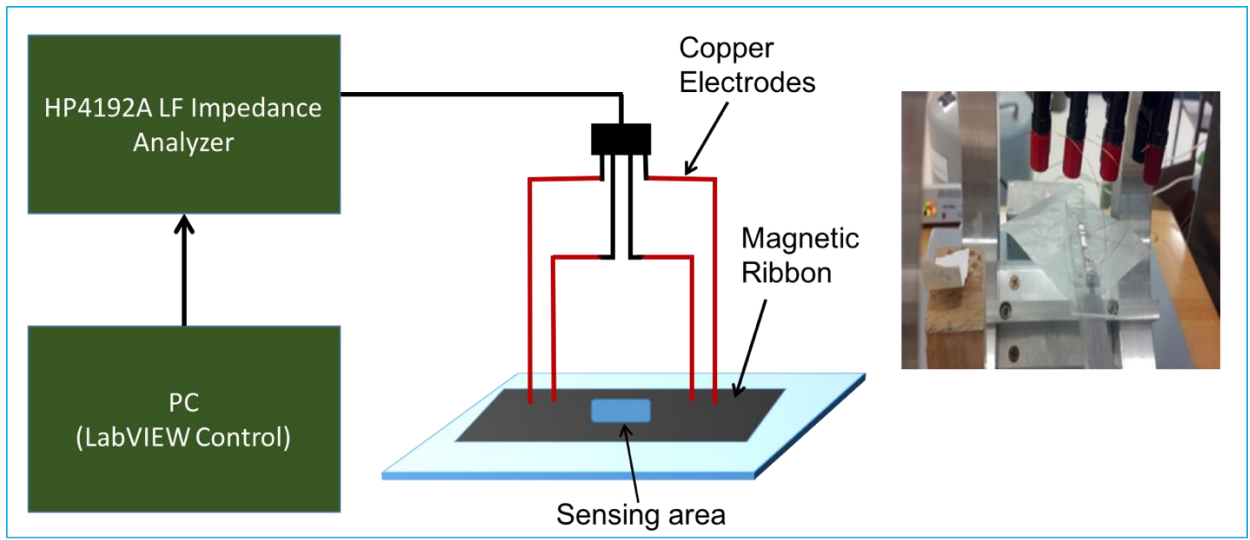

**Figure 2**

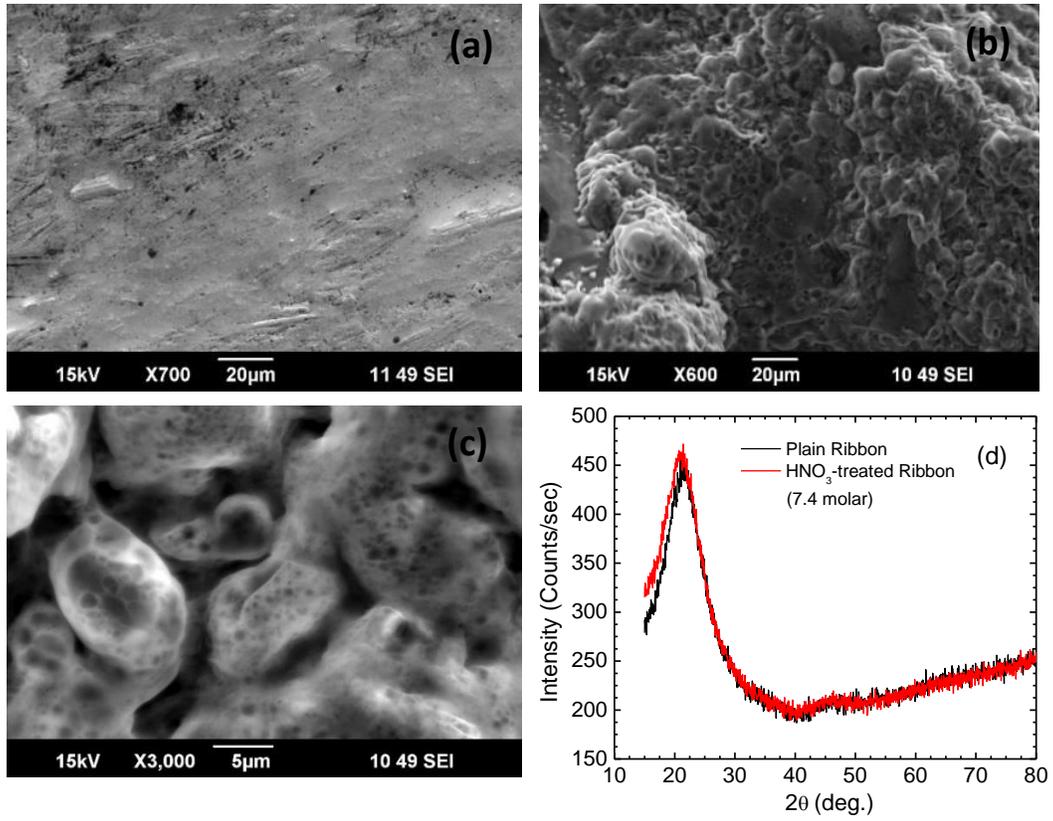

**Figure 3**

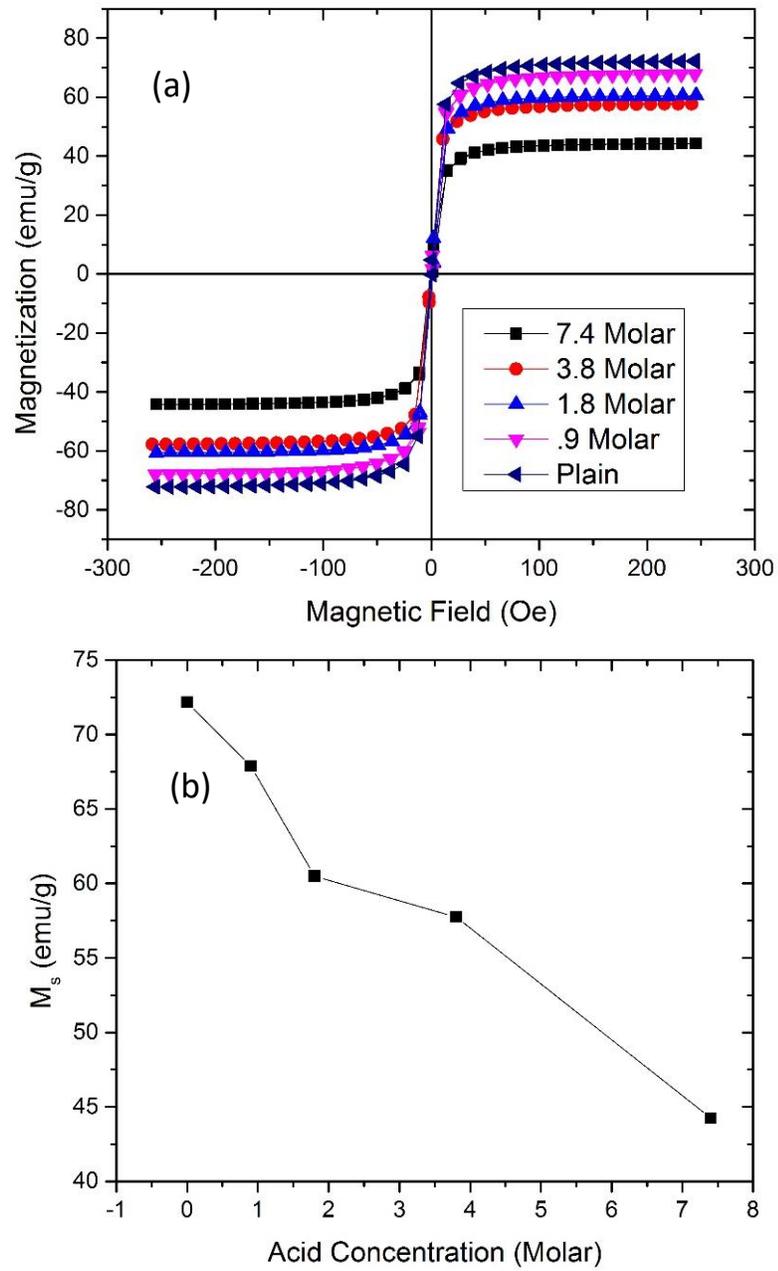

**Figure 4**

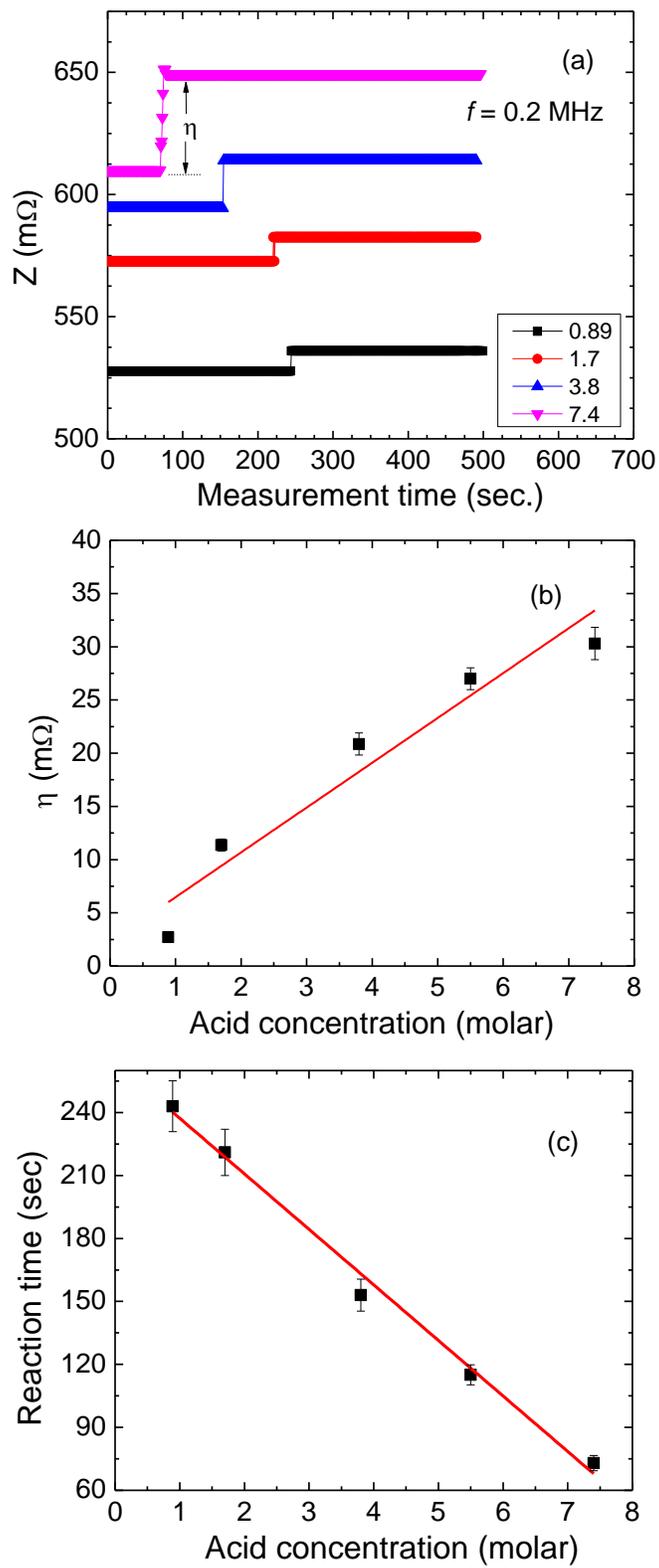

Figure 5

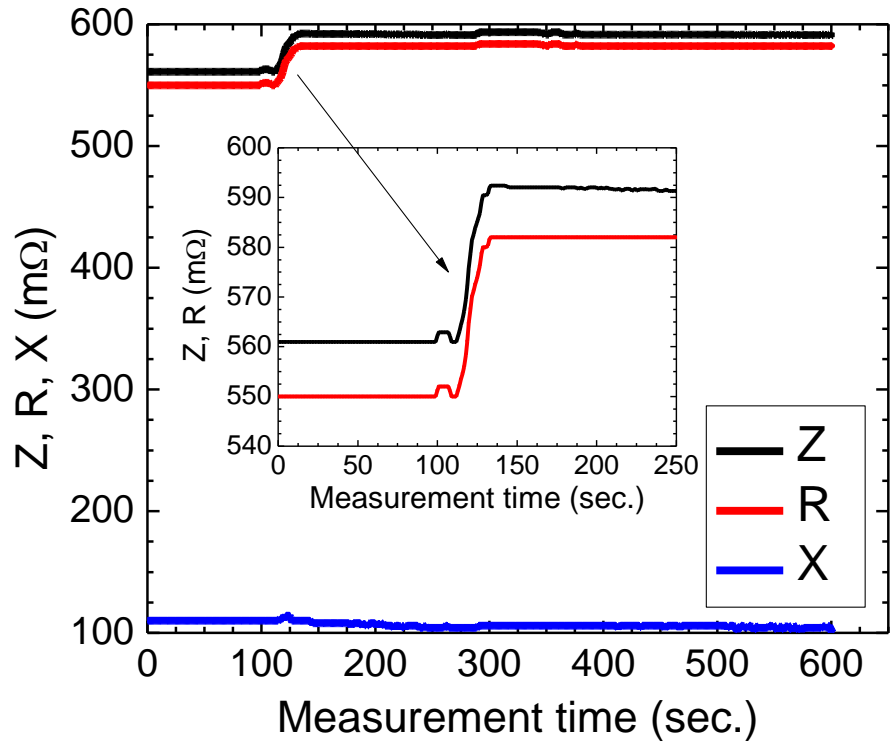